\documentstyle[aps,prb,preprint]{revtex}
\begin{document}
\title{A two band model for Superconductivity:\\
Probing interband pair formation}
\author{R. E. Lagos$^{a)}$ and G. G. Cabrera$^{b)}$}
\address{$^{a)}$ Departamento de F\'{\i }sica, IGCE\\
Universidade Estadual Paulista (UNESP) \\
CP. $178$, $13500$-$970$ Rio Claro, SP, Brazil. \\
$^{b)}$ Instituto de F\'{\i }sica `{\it Gleb Wataghin'} \\
Universidade Estadual de Campinas (UNICAMP)\\
CP. 6165 Campinas, SP 13083-970 Brazil.}
\date{\today}
\maketitle

\begin{abstract}
We propose a two band model for superconductivity. It turns out that the
simplest nontrivial case considers solely interband scattering, and both
bands can be modeled as symmetric (around the Fermi level) and flat, thus
each band is completely characterized by its half-band width $W_{n}$
(n=1,2). A useful dimensionless parameter is $\delta $, proportional to $%
W_{2}-W_{1}$. The case $\delta =0$ retrieves the conventional BCS model. We
probe the specific heat, the ratio gap over critical temperature, the
thermodynamic critical field and tunneling conductance as functions of $%
\delta $ and temperature (from zero to $T_{c}$). We compare our results with
experimental results for $MgB_{2}$ and good quantitative agreement is
obtained, indicating the relevance of interband coupling. Work in progress
also considers the inclusion of band hybridization and general interband as
well as intra-band scattering mechanisms.
\end{abstract}


\narrowtext
\tightenlines

\section{Introduction}

Magnesium Diboride ($MgB_{2}$) appears to be a rather ``unconventional''
conventional superconductor \cite{rev1,rev2}. Two band effects observed as
deviations of conventional BCS\ include: anomalous specific heat \cite{exp1}
and two gaps features (including double peaked tunneling conductance
spectra) \cite{exp2,exp3,exp4,exp5,exp6,exp7,exp8}. The superconductive
mechanism, nevertheless seems to be conventional phonon BCS-like \cite{exp9}%
. In this short communication we present a two band model based on the
classical work by Suhl et al. \cite{suhl} and on an extension of the latter
applied to high $T_{c}$ compounds \cite{lagos}. We mention other multiband
models in the literature \cite
{theo1,theo2,theo3,theo4,theo5,theo6,theo7,theo8}, and some calculations and
fittings within a multiband and strong coupling context include Ref. %
\onlinecite{fit1,fit2,fit3,fit4}. In section II we introduce a two band
model \cite{suhl,lagos} and within the usual BCS scheme we compute the mean
field expressions for the free energy, entropy, critical field, conductance,
and the selfconsistent equations for the gaps functions. In particular we
consider the simplest case: solely interband pairing coupling via phonons.
In section III we compare our simple model with some experimental results
for the case of $MgB_{2}$ \cite{banda,rev1,phonon}, indicating that the
interband pairing mechanism is somehow relevant. Finally in section IV we
present some concluding remarks and future work.

\vspace{0.5cm}

\section{The two band model}

Our model follows Ref. \onlinecite{suhl,lagos}, with the Hamiltonian

\begin{equation}
H=\sum_{{\bf k,}m}E_{k,m}\left( c_{k,m}^{\dagger
}c_{k,m}-c_{-k,m}c_{-k,m}^{\dagger }\right) -\frac{1}{N}\sum_{{\bf kq}%
,m}V_{n,m}c_{k,n}^{\dagger }c_{-k,n}^{\dagger }c_{-q,m}c_{q,m}  \label{ham}
\end{equation}

\noindent where the $c_{k}^{\dagger }$'s are the usual creation operators, $%
E_{k,m}$ are the bands dispersion ($m=1,2$), $V_{n,m}$ are the positive
pairing coefficients $(V_{12}=V_{21}$ and $D=V_{11}V_{22}-V_{12}^{2}\neq 0)$%
. We have defined $k=({\bf k},\uparrow ),$ $-k=(-{\bf k},\downarrow )$, $N$
is the number of sites and the last summation is with the usual energy
cutoff $\omega _{D}$ . The order parameters $\Delta _{n}$ are defined as the
expectation values

\[
\Delta _{n}=\frac{1}{N}\sum_{{\bf k},m}V_{n,m}\left\langle c_{k,m}^{\dagger
}c_{-k,m}^{\dagger }\right\rangle 
\]

The effective Hamiltonian is given by (within the Hartree Fock scheme for
anomalous pairing , see Ref. \onlinecite{lagos})

\[
H_{\mbox{eff}}=NE_{0}+\sum_{k,m}\Psi _{k,m}^{\dagger }\left( E_{k,m}\sigma
_{z}-\Delta _{m}\sigma _{x}\right) \Psi _{k{\bf ,}m} 
\]

\noindent where

\[
E_{0}=\frac{1}{D}\left( V_{22}\Delta _{1}^{2}+V_{11}\Delta
_{2}^{2}-2V_{12}\Delta _{1}\Delta _{2}\right) ,\hspace{0.5cm}\Psi _{k{\bf ,}%
m}\equiv \left( 
\begin{array}{c}
c_{k{\bf ,}m}^{{}} \\ 
c_{-k,m}^{\dagger }
\end{array}
\right) 
\]

\noindent and $\sigma _{x}$, $\sigma _{z}$ are the usual Pauli matrices$.$
The free energy per site $F$ is given by

\[
\exp \left( -\beta NF\right) =\mbox{Tr}\exp \left( -\beta H_{\mbox{eff}%
}\right) 
\]

\[
F=E_{0}+\frac{T}{N}\sum_{k,m}\ln f_{k,m}(1-f_{k,m}) 
\]

\noindent where $f(\omega )=$ $\left( \exp (\beta \omega )+1\right) ^{-1}$, $%
\omega _{k,m}=\sqrt{E_{k,m}^{2}+\Delta _{m}^{2}}$ and $f_{k,m}=f(\omega
_{k,m}).$ The relative free energy $\delta F=F-F(\Delta _{1}=\Delta _{2}=0)$%
, the thermodynamic critical field $H_{c}$, entropy (per site) and specific
heat are given, respectively by

\begin{equation}
\delta F(T)=E_{0}-\frac{T}{N}\sum_{k,m}\ln \frac{\left( 1+\cosh \beta \omega
_{k,m}\right) }{\left( 1+\cosh \beta E_{k,m}\right) }=-\frac{1}{8\pi }%
H_{c}^{2}  \label{obs1}
\end{equation}

\begin{equation}
S=-\frac{2}{N}\sum_{k,m}\left( (1-f_{k,m})\ln (1-f_{k,m})+f_{k,m}\ln
f_{k,m}\right)  \label{obs2}
\end{equation}

\begin{equation}
C_{V}=T\left( \frac{\partial S}{\partial T}\right) _{V}=\frac{2\beta ^{2}}{N}%
\sum_{k,m}f_{k,m}(1-f_{k,m})\left( \omega _{k,m}^{2}+\frac{1}{2}\beta \frac{%
\partial \Delta _{m}^{2}}{\partial \beta }\right)  \label{obs3}
\end{equation}

The condensation energy is given by

\begin{equation}
\delta F(T=0)=W_{C}=E_{0}-\frac{1}{N}\sum_{k,m}\left( \omega
_{k,m}-E_{k,m}\right)  \label{cond}
\end{equation}

\noindent and the superconductor- normal tunneling differential conductance
(conveniently scaled) is defined by

\begin{equation}
G(V)=-\sum_{m}\int d\varepsilon \rho _{m,S}(\varepsilon )\frac{\partial
f(\varepsilon +V)}{\partial \varepsilon }  \label{conduc}
\end{equation}

\[
\rho _{m,S}(\varepsilon )=\rho _{m}\left( \mbox{sign}(\varepsilon )\sqrt{%
\varepsilon _{{}}^{2}-\Delta _{m}^{2}}\right) \mbox{Real}\left( \sqrt{\frac{%
\left( \varepsilon +i\Gamma \right) ^{2}}{\left( \varepsilon +i\Gamma
\right) ^{2}-\Delta _{m}^{2}}}\right) ,\hspace{0.5cm}\Gamma \rightarrow
0^{+} 
\]

Minimization of the free energy with respect to the gaps functions$,$
\noindent yields a coupled nonlinear system of integral equations for the
gaps, to be solved selfconsistently, and given by

\begin{eqnarray}
\left( V_{22}-DR_{1}(\Delta _{1},T)\right) \Delta _{1}-V_{12}\Delta _{2} &=&0
\nonumber \\
&&  \label{self} \\
-V_{12}\Delta _{1}+\left( V_{11}-DR_{2}(\Delta _{2},T)\right) \Delta _{2}
&=&0  \nonumber
\end{eqnarray}

\noindent where

\[
R_{m}(\Delta _{m},T)=\int_{-\omega _{D}}^{+\omega _{D}}d\varepsilon \rho
_{m}(\varepsilon )S\left( \sqrt{\varepsilon ^{2}+\Delta _{m}^{2}}\right) ,%
\hspace{0.3cm}S(x)=\frac{1}{2x}\tanh \left( \frac{x}{2T}\right) 
\]

\noindent and with $\rho _{m}(\varepsilon )$ the density of states
associated to the respective band. \noindent The transition temperature is
the highest temperature $T_{c}=\beta _{c}^{-1},$ solution of

\[
\left( V_{22}-DR_{1}(0,T_{c})\right) \left( V_{11}-DR_{2}(0,T_{c})\right)
=V_{12}^{2} 
\]

\vspace{0.5cm}

\section{Results}

We compute the observables presented in the previous section. In particular
we consider only interband scattering $V_{11}=V_{22}=0,$ $V_{12}=\lambda $,
the simplest relevant case \cite{suhl,lagos}. We consider two flat symmetric
bands, with $\rho _{m}(\varepsilon )\equiv \rho _{m}(0)=\rho _{m}$.

The gaps equations (\ref{self}) now read

\[
\Delta _{m}=\lambda \rho _{n}\Delta _{n}R(\Delta _{n},T),\hspace{0.5cm}n\neq
m=1,2 
\]

\[
R(\Delta ,T)=\int_{0}^{\omega _{D}}\frac{d\varepsilon }{\sqrt{\varepsilon
^{2}+\Delta ^{2}}}\tanh \left( \frac{\beta }{2}\sqrt{\varepsilon ^{2}+\Delta
^{2}}\right) 
\]

{\bf At zero temperature} the gaps equations are given by (in convenient
units)

\begin{eqnarray}
\phi _{1} &=&\frac{2\Delta _{1}(T=0)}{3.53T_{c}}=\exp \frac{1}{\xi }\left(
1-a\right)  \nonumber \\
&&  \label{gaps} \\
\phi _{2} &=&\frac{2\Delta _{1}(T=0)}{3.53T_{c}}=\exp \frac{1}{\xi }\left( 1-%
\frac{1}{a}\right)  \nonumber
\end{eqnarray}

\noindent where $\xi ^{2}=\lambda ^{2}\rho _{1}\rho _{2}$ and $a$ satisfies
a selfconsistent equation. An excellent approximate solution is given by

\[
\ln a=\frac{\xi \theta }{2-\theta },\hspace{0.5cm}\theta =\ln \sqrt{\frac{%
1+\delta }{1-\delta }} 
\]

\noindent with

\[
-1<\delta =\frac{\rho _{1}-\rho _{2}}{\rho _{1}+\rho _{2}}<1 
\]

Notice that all the above mentioned observables will yield the standard BCS
expressions \cite{crisan} in the limit $\delta =0$.

The critical temperature is given by $T_{c}=1.13\omega _{D}\exp (-\xi
^{-1}). $

We label the bands such that $\delta >0.$ If we consider $MgB_{2}$, from
Ref. \onlinecite{rev1,phonon} we have $T_{c}\simeq 40$ $^{0}K$, $\omega
_{D}\simeq 800^{0}K$ yielding $\xi \simeq 0.32$. From Ref. \onlinecite{banda}
we approximate $W_{1}\approx 5.6eV$ $(\rho _{1}\approx 0.179eV^{-1}),$ $%
W_{2}\approx 14.eV$ $(\rho _{2}\approx 0.071eV^{-1})$ yielding $\delta
\approx 0.432$.

In Figure 1 we plot the normalized gaps $\phi _{m}$ at zero temperature, Eq.(%
\ref{gaps}), and (minus) the condensation energy $-W_{c}$ Eq.(\ref{cond}),
both as function of $\delta .$ The condensation energy is normalized to the
BCS reference state $i.e.$ $W_{c}=\delta F(\delta ,T=0)/\delta F(\delta
=0,T=0)$ (see Eq.\ref{obs1}). The chosen normalization yields the standard
BCS (weak coupling) value of unity for the gaps and the condensation energy.
As $\delta $ is varied away from zero the condensation energy is less than
the standard BCS. One gap will depart from weak to `a medium coupling
regime' ($\phi _{2}>1$), conversely the other gap will dive towards `a less
than weak coupling regime' ($\phi _{1}<1$), with the geometrical average $%
\sqrt{\phi _{1}(\delta )\phi _{2}(\delta )}\equiv 1$ always in the standard
weak coupling regime. These features seem consistent as we fit the parameter 
$\delta $ with experimental data \cite{rev1}-\cite{exp8}.

In order to solve for the gaps, Eq.(\ref{self} ), we can use the available
low temperature and near the critical temperature expansions \cite{crisan}.
These allow us to nicely interpolate, for the full temperature regime $0\leq
\tau =T/T_{c}\leq 1$. Once this is done we can readily compute the specific
heat, Eq.(\ref{obs3}), entropy, Eq.(\ref{obs2}), and the thermodynamic
critical field, Eq.(\ref{obs1}).

In Figure 2 we plot the specific heat $C_{V}$ (normalized to the normal
state value at $T_{c}$) versus the temperature $\tau $ for several values of 
$\delta .$ The standard BCS result is represented by the curve $\delta =0.$
The anomalous behavior of $C_{V}$ consists in going under the BCS\ value in
the region $0.5<\tau <1$, and going over the BCS value in the region $0<\tau
<0.5.$ This feature is in very good agreement with Ref. \onlinecite{exp1}.
In Figure 3 we plot the entropy $S$ (normalized to the normal state value at 
$T_{c}$) versus the temperature $\tau $ for several values of $\delta .$ The
standard BCS result is represented by the curve $\delta =0.$ As $\delta $
departs from zero (bands are less `identical') the system increases its
entropy. In Figure 4 we plot the thermodynamic critical field (normalized to
the reference state $\delta =0$, $T=0$) versus the temperature $\tau $ for
several values of $\delta .$ The standard BCS result is again represented by
the curve $\delta =0.$ As $\delta $ increases the critical field is reduced
when compared to the BCS value. This is in agreement with experimental
results \cite{exp1}.

In Figure 5 we plot the conductance, Eq.(\ref{conduc}) versus applied
voltage, for a fixed value of $\delta =0.5$, and for several temperatures,
and where a small dispersion is included, $\Gamma =0.1$ meV \cite{fit1}.The
double peaked form is in very good agreement with observations (see for
example Ref. \onlinecite{exp5}).

\vspace{0.5cm}

\section{Concluding Remarks}

We presented the simplest relevant two band model for superconductivity,
based on a standard BCS-like pairing mechanism. We computed the gaps
equations at zero temperature. Also the specific heat, entropy, critical
field and conductance as function of temperature. We considered the simplest
interband scattering mechanism (one pairing parameter) and two planar
symmetrical bands (one parameter band model). Our results seems to be in
very good agreement with some experimental results on the compound $MgB_{2}$%
, indicating that interband pairing is somehow relevant for this compound.
These results are being investigated further. Work in progress incorporates
intraband pairing mechanisms, an hybrid-like interband pairing mechanism 
\cite{lagos}, absent in most of the theoretical models, and a more involved
band structure.

\newpage

\begin{center}
{\large {\bf Figure Captions}}
\end{center}

\vspace{0.5cm}

\noindent {\bf Figure 1}: Gaps $\phi _{m}$ at zero temperature and (minus)
the condensation energy $-W_{c}$ versus $\delta .$ Convenient units $\phi
_{m}=2\Delta _{m}^{0}/3.53T_{c}$ and $W_{c}=\delta F(\delta ,T=0)/\delta
F(\delta =0,T=0)$. See text.

\vspace{0.5cm}

\noindent {\bf Figure 2}: Specific heat $C_{V}$ versus $\tau =T/T_{c}$ for
several $\delta $ values, normalized to the normal state specific heat at $%
T_{c}$: $C_{n}(T_{c})=4\pi ^{2}(\rho _{1}+\rho _{2})/6.$ See text.

\vspace{0.5cm}

\noindent {\bf Figure 3}: Entropy $S$ versus $\tau =T/T_{c}$ for several $%
\delta $ values; normalized to the normal state entropy at $T_{c}$: $%
S_{n}(T_{c})=4\pi ^{2}(\rho _{1}+\rho _{2})T_{c}/6.$ See text.

\vspace{0.5cm}

\noindent {\bf Figure 4}: Thermodynamic critical field $H_{c}$ versus $\tau
=T/T_{c}$ for several $\delta $ values, normalized to the reference (BCS)
state $T=0$, $\delta =0,$ $H_{c}^{2}(\tau )=F(\tau ,\delta )/F(0,0).$ See
text.

\vspace{0.5cm}

\noindent {\bf Figures 5}: Tunneling conductance for several temperatures,
at $\delta =0.5$, versus applied voltage. A small dispersion is included $%
\Gamma =0.1$ meV \cite{fit1}. See text.

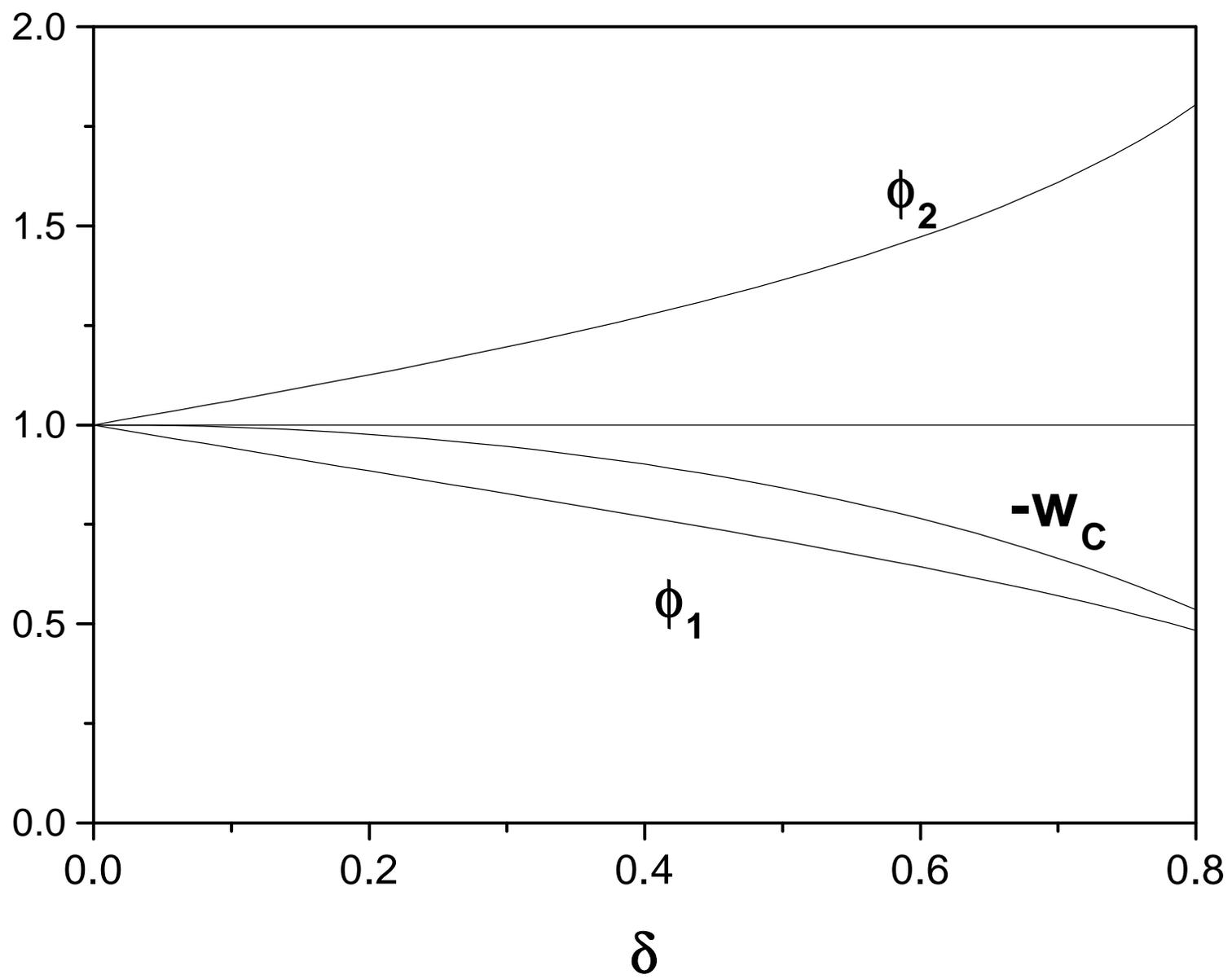
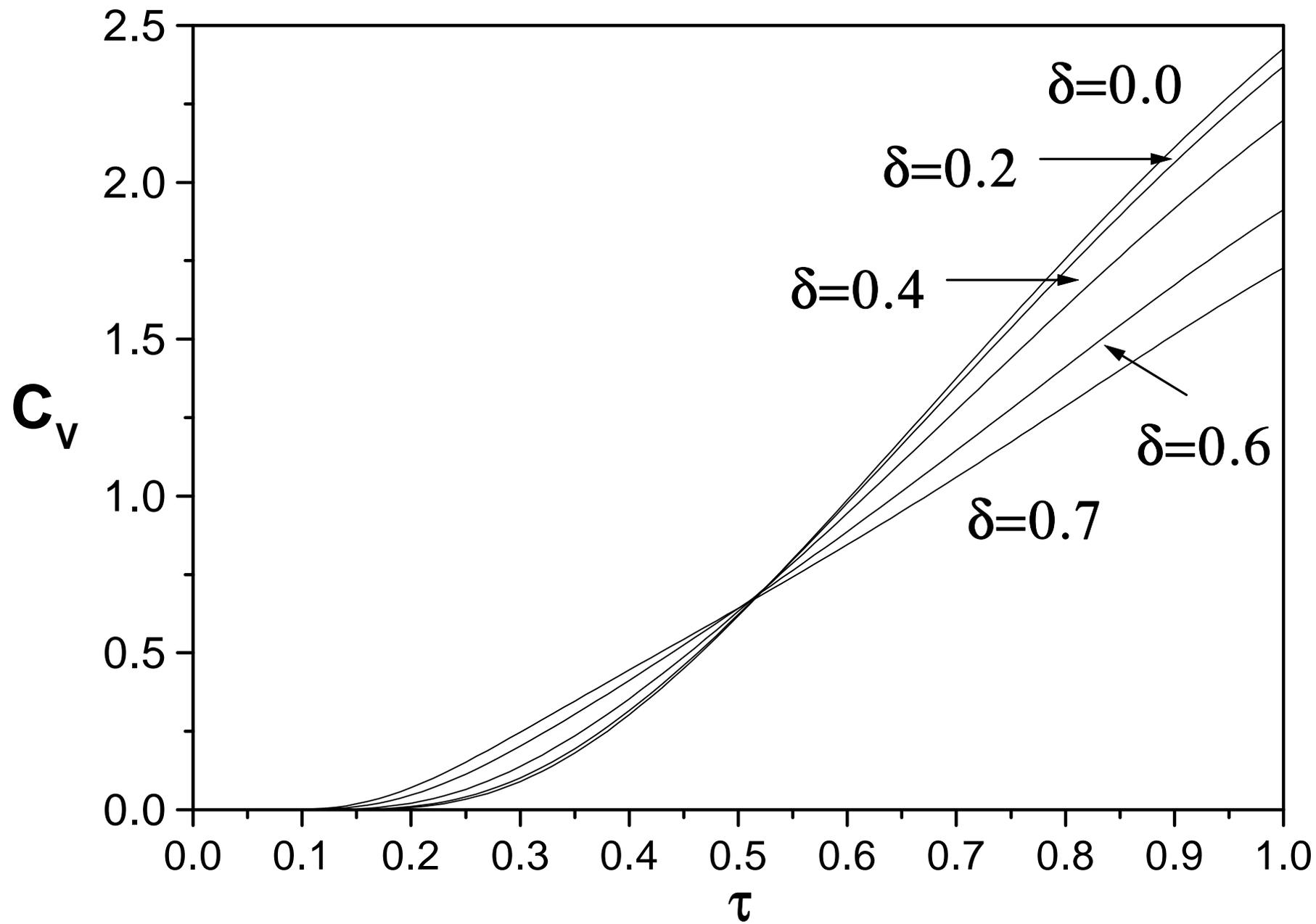
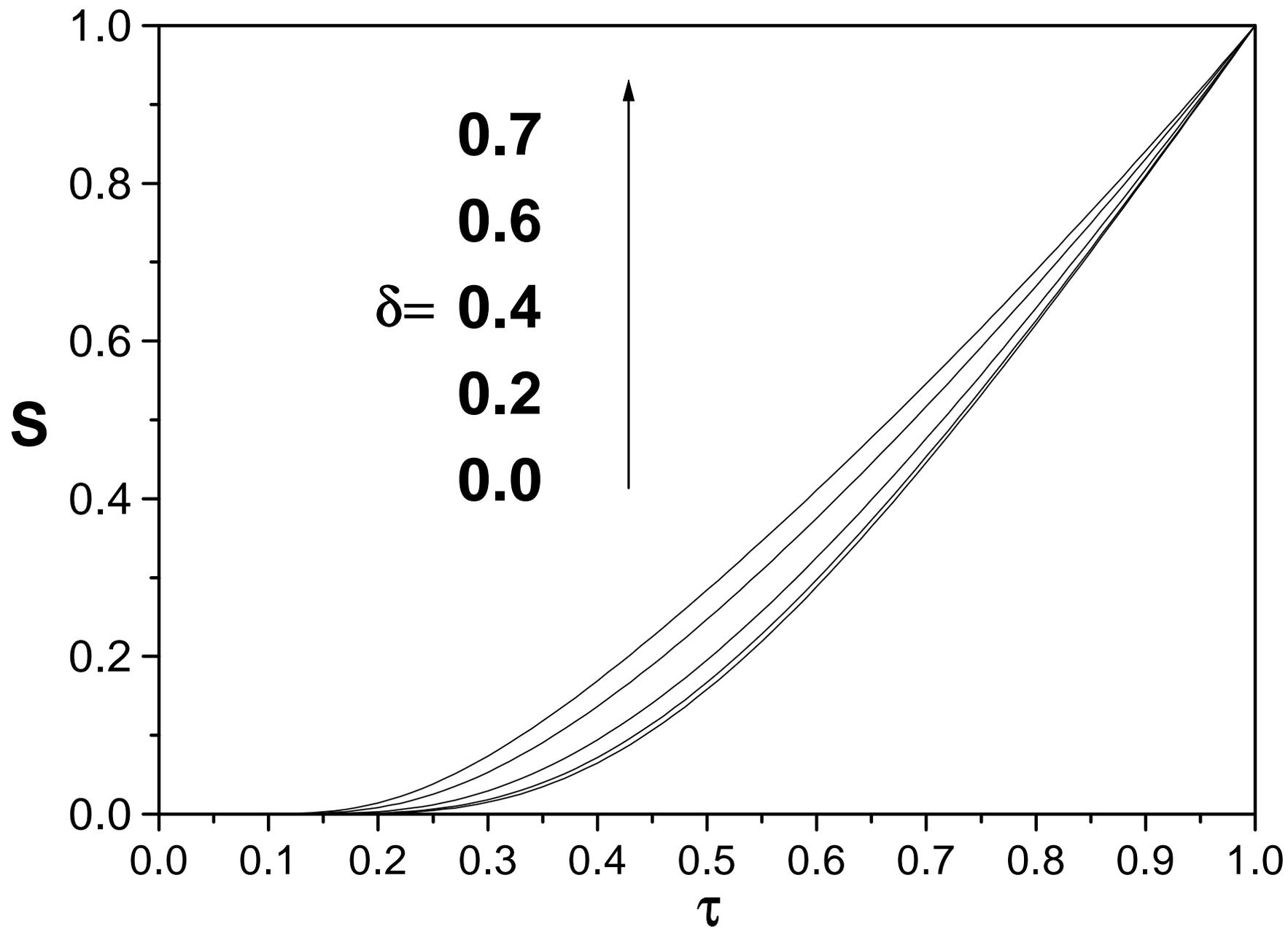
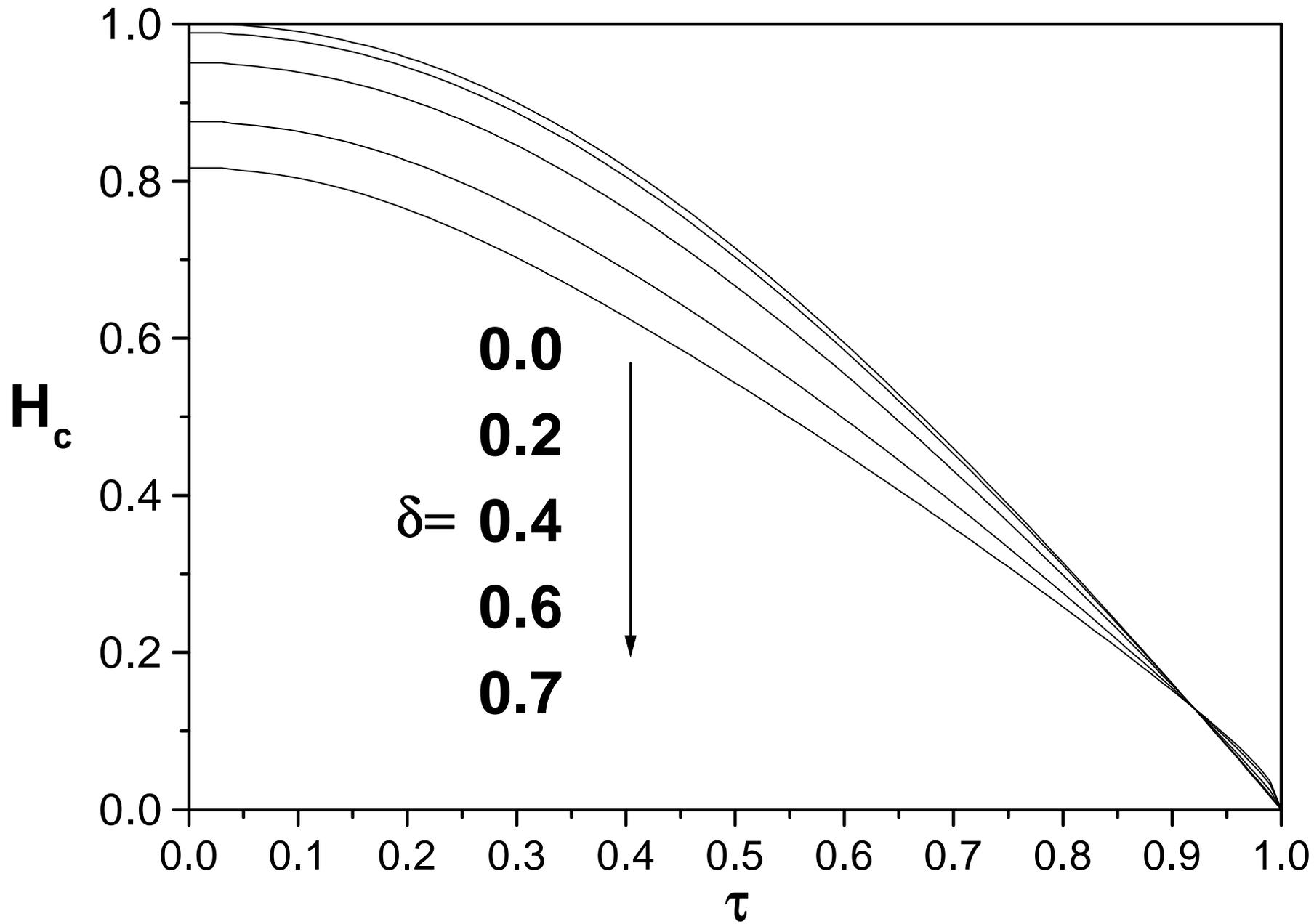
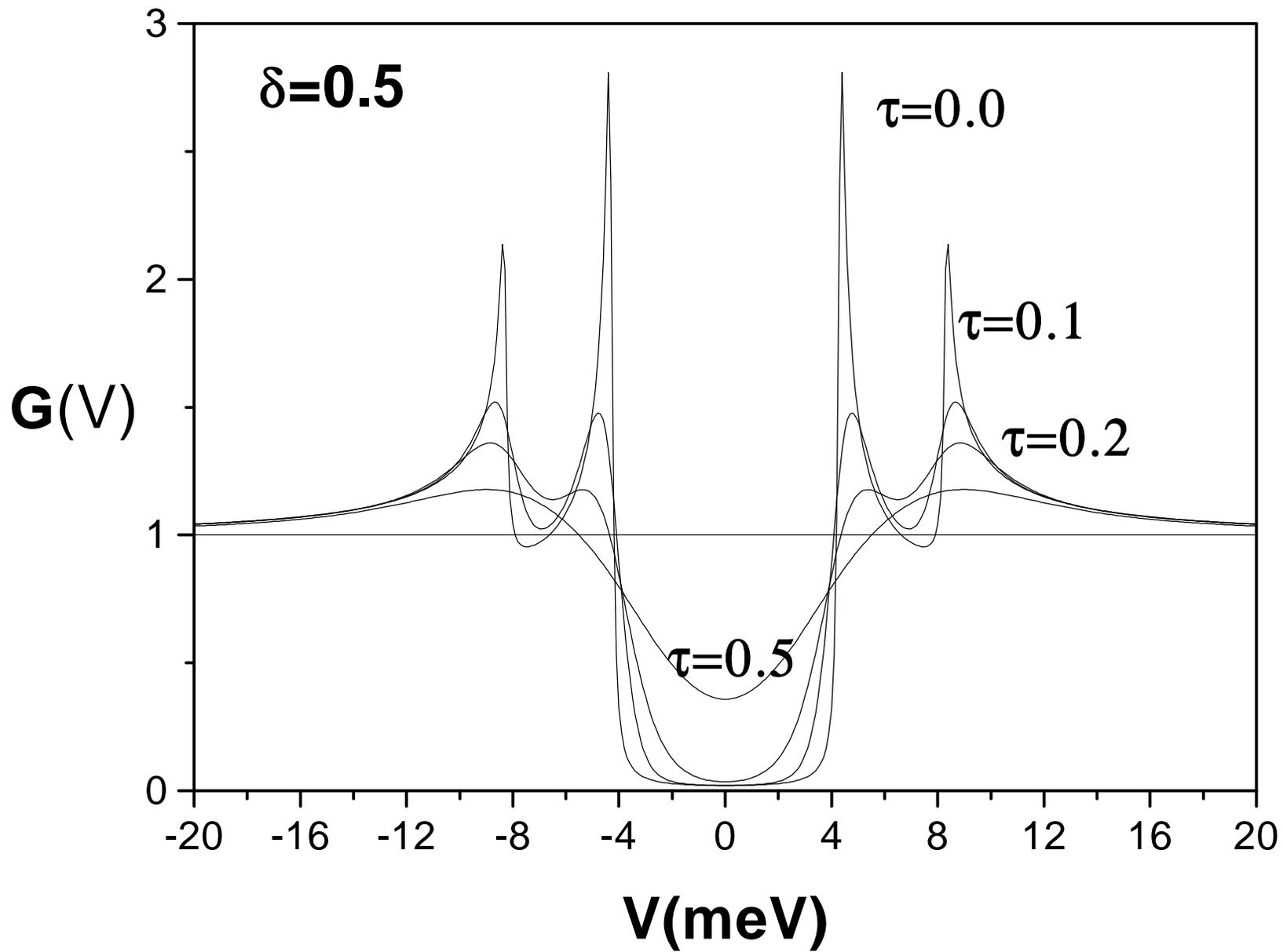
\end{document}